\documentclass[11pt,a4paper]{article}

\usepackage[margin=1in]{geometry}
\usepackage{amsmath,amssymb}
\usepackage{hyperref}
\usepackage{microtype}
\usepackage{graphicx}
\usepackage{booktabs,flafter,placeins}

\hypersetup{
    pdftitle={Dynamics of a Charged Particle on a Conformal Helicoidal Surface},
    pdfauthor={Hassan Hassanabadi, Abdullah Guvendi, Georg Junker},
    colorlinks=true,
    linkcolor=blue,
    citecolor=blue,
    urlcolor=blue
}

\title{Dynamics of a Charged Particle on a Conformal Helicoidal Surface}
\author{Hassan Hassanabadi\textsuperscript{1,2,*},
Abdullah Guvendi\textsuperscript{3,$\dag$} and Georg Junker\textsuperscript{4,$\ddag$}\\[1ex]
\small\textsuperscript{1}\,Department of Physics, Faculty of Science, University of Hradec Kr\'alov\'e, Czechia\\
\small\textsuperscript{2}\,Department of Physics, California State University, Fresno, CA 93740, USA\\
\small\textsuperscript{3}\,Department of Basic Sciences, Erzurum Technical University, Erzurum, T\"urkiye\\
\small\textsuperscript{4}\,Institute for Theoretical Physics, Friedrich-Alexander-Universit\"at\\[-0.15ex]
\small Erlangen-N\"urnberg, 91058 Erlangen, Germany\\[0.5ex]
\small\textsuperscript{*}\,\href{mailto:hassan.hassanabadi@uhk.cz}{hassan.hassanabadi@uhk.cz}\\
\small\textsuperscript{$\dag$}\, \href{mailto:abdullah.guvendi@erzurum.edu.tr}{abdullah.guvendi@erzurum.edu.tr}\\
\small\textsuperscript{$\ddag$}\,Corresponding author, \href{mailto:georg.junker@fau.de}{georg.junker@fau.de}
}
\date{}

\begin{document}
\maketitle

\begin{abstract}
We study the classical and quantum motion of a charged particle on a conformal helicoidal surface embedded in four-dimensional Euclidean space. A uniform ambient magnetic field pulls back to a nonuniform surface field. The cyclic longitudinal coordinate permits an exact transverse reduction. The ordinary helicoid in the same axial field shares the effective potential and the turning points, while the conformal metric increases the physical period at matched energy and longitudinal momentum. We classify the single- and double-well regimes and obtain the exact transverse periods in the two double-well sectors. The periods diverge logarithmically at the separatrix, with twice the leading coefficient above the barrier as below it. The zero-field limit allows transverse escape, and the weak-twist limit is locally free but nonuniform across the surface. Intrinsic quantization gives an exact weighted transverse eigenvalue problem. Fixed-momentum eigenvalues test the harmonic approximation in a central well and resolve the tunneling splittings in a double well.
\end{abstract}

\section{Introduction}
Dynamics on two-dimensional surfaces is a core topic for a vast range of scientific disciplines such as biology, technical engineering and material science but also attracts much interest in mathematics as well as in classical and quantum physics \cite{Steeneken2024}. With new materials like graphene, two-dimensional dynamics is nowadays realisable for almost arbitrary surface shapes \cite{Politano2026}. The dynamics of a particle on a curved surface depends on both the induced metric and the way an external field couples to it. In particular, a helicoidal surface is widely used in those fields mentioned before.

With external axial magnetic field as will be considered here, helicoidal symmetry makes the longitudinal coordinate cyclic. Eliminating it at fixed momentum is an instance of what is called Routh reduction in the literature \cite{MarsdenRatiuScheurle2000}. Geodesics on screw-symmetric tubes and integrable magnetic flows on the two-torus provide examples of such symmetry-based integration \cite{JantzenHelical,AsselleBenedetti2021}. Curvature-induced localization and spectral changes have been studied on helical nanotubes and nanoribbons \cite{AtanasovDandoloff2008,Balakrishnan2025}. On helicoidal strips, geometric potentials give rise to transverse localization and charge separation \cite{DandoloffTruong2004,Atanasov2009}. Studies of helicoidal graphene nanoribbons address both geometry-induced potentials and the response to an axial magnetic field \cite{AtanasovSaxena2015,Soares2023}. Effective operators with curvature-dependent terms have also been derived for thin strips on ruled surfaces in higher dimensions~\cite{KrejcirikZahradova2020}.

In an earlier paper Guvendi \emph{et al.} derived a classical Hamiltonian for the ordinary helicoid in a uniform axial field \cite{Guvendi2026}, and related work examines curvature and gauge coupling in a helicoidal two-body system \cite{GuvendiHassanabadi2026}. Here we will use the same projected helicoid and ambient magnetic two-form as in the one-particle model. Whereas in \cite{Guvendi2026} the helicoid was embedded in a 3-dimensional Euclidean space, here we pursue an embedding in $\mathbb{R}^4$. The common longitudinal metric and gauge pullback give the same transverse effective potential at fixed longitudinal momentum. However, the change in induced metric affects the transverse inertia, physical period, curvature, and magnetic flux per unit area.

For quantization we use the intrinsic magnetic Laplace--Beltrami Hamiltonian. In thin-layer models, physical confinement normal to the surface can induce geometric terms as discussed by Jensen and Koppe \cite{JensenKoppe1971} as well as by da Costa \cite{daCosta1981,daCosta1982}. Strong-confinement and adiabatic analyses clarify how these terms arise~\cite{FroeseHerbst2001,WachsmuthTeufel2010}, and transverse-mode degeneracies can induce effective gauge connections~\cite{Mitchell2001,SchusterJaffe2003}. For charged particles, gauge-consistent treatments of the normal dynamics give the effective electromagnetic coupling on the surface~\cite{FerrariCuoghi2008,OrtixVanDenBrink2011}.

The Landau problem on the hyperbolic plane is an example of a magnetic spectrum shaped by the surface metric~\cite{Comtet1987}. Fixing longitudinal momentum yields a family of transverse spectral problems, as in planar models with a magnetic field that varies along one coordinate~\cite{Iwatsuka1985,DombrowskiHislopSoccorsi2014}. This reduction lets us compare the exact transverse eigenvalues with semiclassical quantization and double-well tunneling formulas~\cite{BerryMount1972,Simon1984,Garg2000}.

Our paper is organised as follows. In the next section, we first construct the conformal embedding and compute the induced curvature. Section 3 introduces a uniform ambient magnetic two-form and the surface pullback and flux are presented.
We then study the classical dynamics of a charged test mass. For different parameters of our model the corresponding effective potential exhibits either a double-well or single-well structure. Accordingly we classify the transverse orbits, obtain the periods on both sides of the separatrix, and study the zero-field and weak-twist limits. In section 5 we consider the corresponding quantum dynamics generated by the magnetic Laplace-Beltrami operator and calculate within a semi-classical approximation the leading-order WKB eigenvalues and the related tunneling splittings. Section 6 compares the semiclassical predictions with numerical results of the exact transverse quantum dynamics, followed by a summary and final discussion.

\section{Conformal Helicoidal Embedding and Curvature}

Let $(x_{\rm E},y_{\rm E},z_{\rm E},\xi)$ denote the Cartesian coordinates in Euclidean space $\mathbb R^4$. Further, let $(u,v)\in\mathbb R^2$ be  surface coordinates and $\omega>0$ be the twist rate. The embedding, we will be considering throughout this paper, is then given by
\begin{equation}\label{embedding}
	X(u,v)=\bigl(v,u\cos(\omega v),u\sin(\omega v),\tfrac12\omega u^2\bigr).
\end{equation}
For fixed $u\ne0$, the projection into the first three coordinates is a circular helix of radius $|u|$ with axial advance $2\pi/\omega$ per turn. The fourth coordinate is constant along each such helix.

The tangent vectors  $X_u=\frac{\partial X}{\partial u}$ and $X_v=\frac{\partial X}{\partial v}$ are orthogonal and both have the squared norm
\begin{equation}\label{Omega2}
  \Omega^2(u)=1+\omega^2u^2.
\end{equation}
Their Gram matrix gives the induced conformal metric
\begin{equation}\label{ds2}
	\mathrm{d}s^2=\Omega^2(u)(\mathrm{d}u^2+\mathrm{d}v^2)
\end{equation}
and \eqref{Omega2} plays the role of a conformal factor.
In these coordinates the metric determinant is $g=\Omega^4$ and the
area element reads $\Omega^2\,\mathrm{d}u\,\mathrm{d}v$. Let us compare this with the ordinary
helicoid discussed in \cite{Guvendi2026} exhibiting in the same coordinates a metric of the form
\begin{equation}\label{ds2hel}
  \mathrm{d}s_{\rm hel}^2=\mathrm{d}u^2+\Omega^2(u)\,\mathrm{d}v^2.
\end{equation}
The added fourth coordinate contributes $(\partial_u\xi)^2=\omega^2u^2$ only to the transverse
metric coefficient. In the weak-twist limit $\omega\to0$ the embedding
approaches the plane $(v,u,0,0)$ on bounded coordinate patches.

\begin{figure}[!t]
	\centering
	\includegraphics[width=\columnwidth]{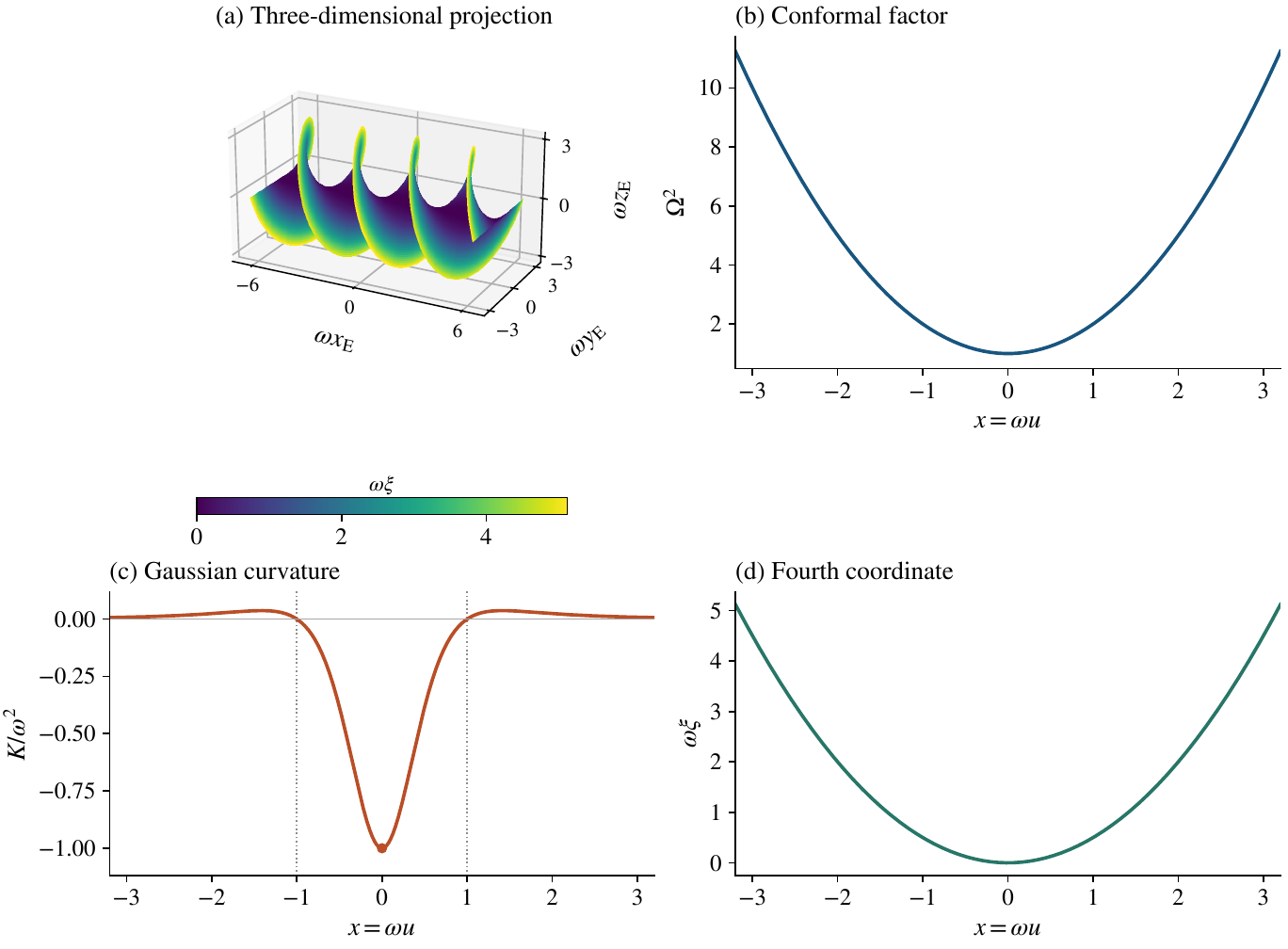}
	\caption{Geometry of the conformal helicoidal surface.
	(a) Three-dimensional projection with dimensionless axes
	$(\omega x_{\rm E},\omega y_{\rm E},\omega z_{\rm E})$ and color
	representing $\omega\xi$. The patch spans $x=\omega u\in[-3.2,3.2]$
	and $\omega v\in[-2\pi,2\pi]$.
	(b) Conformal factor $\Omega^2=1+x^2$.
	(c) Normalized Gaussian curvature. The marker shows the minimum at
	$x=0$ and the dotted lines locate the zeros at $x=\pm1$.
	(d) Dimensionless fourth coordinate $\omega\xi=x^2/2$.}
	\label{fig:1}
\end{figure}

The Gaussian curvature $K$ of the conformal metric \eqref{ds2} follows directly
from $K=-\Omega^{-2}(\ln\Omega)''$, with primes denoting $u$ derivatives.
Substituting the conformal factor \eqref{Omega2} results in
\begin{equation}\label{curvature}
	K(u)=\frac{\omega^2(\omega^2u^2-1)}{(1+\omega^2u^2)^3}.
\end{equation}
The minimum curvature is reached at $u=0$ with value $K(0)=-\omega^2$. The curvature is negative for
$|u|<1/\omega$, vanishes at $|u|=1/\omega$, and is positive farther out.
For large values $|u|$ it tends to zero as $K(u)\sim 1/(\omega^2u^4)$.
Figure~\ref{fig:1} shows the embedding and the corresponding
dimensionless profiles. Finally we note that the orthonormal tangent frame is represented by $e_u=X_u/\Omega$ and $e_v=X_v/\Omega$ and is completed by the two unit normals
$n_1=(x,\sin \omega v,-\cos \omega v,0)/\Omega$ and $n_2=(0,x \cos \omega v,x \sin \omega v,-1)/\Omega$, where we introduced the dimensionless coordinate $x=u\omega$.

\section{Magnetic Field, Surface Pullback, and Flux}\label{sec:magnetic}
Consider the uniform ambient magnetic two-form
$F_{\rm amb}=B\,\mathrm{d}y_{\rm E}\wedge\mathrm{d}z_{\rm E}$ with constant $B\in\mathbb{R}$. This
corresponds to an axial homogeneous magnetic field $B\hat x_{\rm E}$ in the
projected three-space. The Jacobian of $(y_{\rm E},z_{\rm E})$ with
respect to $(u,v)$ is $\omega u$, so the pullback to the surface is
\begin{equation}\label{Finduced}
	F_\Sigma=B\omega u\,\mathrm{d}u\wedge\mathrm{d}v.
\end{equation}
With the orientation $\mathrm{d}u\wedge\mathrm{d}v$, the field per unit
surface area is $\mathcal B=F_\Sigma(e_u,e_v)$. Dividing the coordinate
coefficient in Eq.~\eqref{Finduced} by the area density gives
\begin{equation}\label{Bintrinsicfinal}
	\mathcal B(u)=\frac{B\omega u}{1+\omega^2u^2} = B\,\frac{x}{1+x^2}.
\end{equation}
The projected vector $B\hat x_{\rm E}$ has tangent component
$(B/\Omega)e_v$ and normal component $(Bx/\Omega)n_1$ in the
four-dimensional frame. The intrinsic field in Eq.~\eqref{Bintrinsicfinal}
is determined by the pullback of the two-form.

\begin{figure}[!htbp]
	\centering
	\includegraphics[width=\columnwidth]{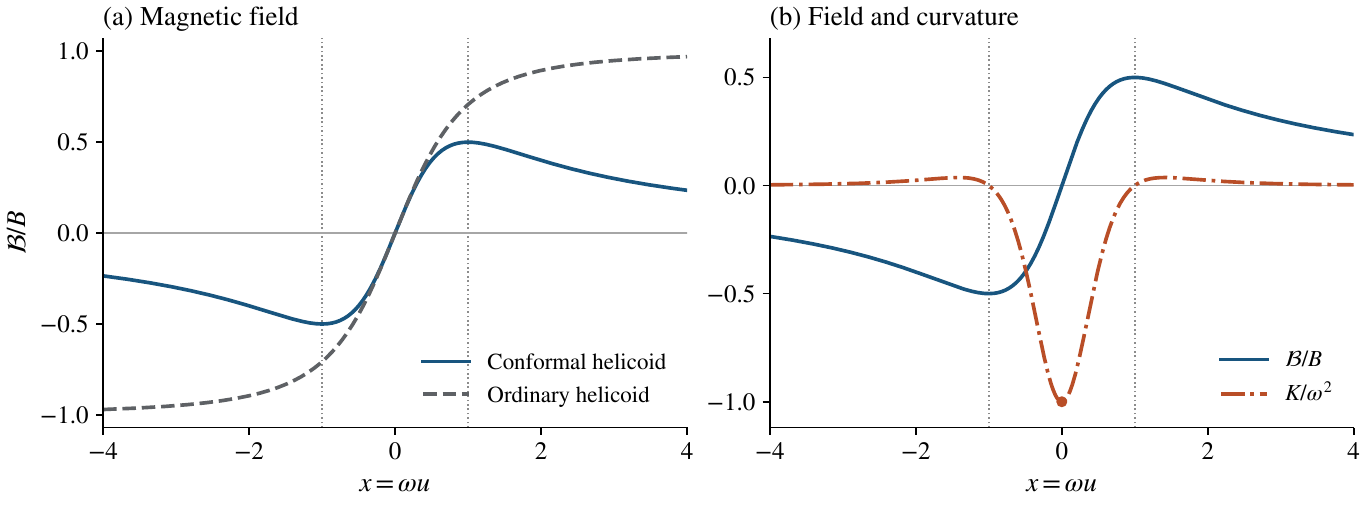}
	\caption{Dimensionless field and curvature profiles.
	(a) Intrinsic magnetic field for the conformal and ordinary
	helicoids in the same axial field. (b) The field and Gaussian
	curvature of the conformal helicoid. The dotted lines mark
	$x=\pm1$, where the extrema of the field coincide with the
	curvature zeros.}
	\label{fig:2}
\end{figure}

Here we note that both embeddings, the conformal as well as the ordinary helicoid, share the same  $F_\Sigma$. On the ordinary helicoid, however, the area density is $\Omega$, giving $\mathcal B_{\rm hel}=Bx/\sqrt{1+x^2}$. Figure~\ref{fig:2} compares the normalized profiles of both embeddings.

The conformal field decays at both transverse ends and reaches
$|\mathcal B|=|B|/2$ at $x=\pm1$, representing the zeros of the Gaussian curvature.
A convenient gauge is obtained by pulling back the symmetric ambient
potential $\tfrac B2(y_{\rm E}\,\mathrm{d}z_{\rm E}
-z_{\rm E}\,\mathrm{d}y_{\rm E})$. On either surface it becomes
\begin{equation}\label{Av}
	A=\tfrac12 B\omega u^2\,\mathrm{d}v.
\end{equation}
In this gauge the vector potential has a vanishing $u$-component $A_u=0$ and its $v$-component is given by $A_v=\tfrac12 B\omega u^2$. Integrating
$F_\Sigma=\mathrm{d}A$ over a rectangle
$[u_1,u_2]\times[v_1,v_2]$ gives the oriented flux
\begin{equation}\label{flux}
	\Phi_\Sigma=\tfrac12 B\omega(v_2-v_1)(u_2^2-u_1^2).
\end{equation}
For one turn and $0\le u\le R$, the flux is $\pi B R^2$.
It vanishes on symmetric interval $-R\le u\le R$ by the oddness of the two-form
coefficient. At fixed finite $u$, both $\mathcal B$ and $A_v$ vanish
as $\omega\to0$.

\FloatBarrier
\section{The Classical Dynamics}

\subsection{Hamiltonian Reduction}
For a particle of mass $m$ and charge $q$, a minimal coupling to the vector potential \eqref{Av} is achieved via the additional term $qA_v\dot v$ to the kinetic Lagrangian. With
$\beta=qB\omega/2$, the reduced-coordinate Lagrangian reads
\begin{equation}\label{eq:lagrangian}
	L=\tfrac12 m\Omega^2(u)(\dot u^2+\dot v^2)+\beta u^2\dot v.
\end{equation}
The momentum $p_u=\partial L/\partial \dot{u}$ equals $m\Omega^2\dot u$, and the cyclic coordinate
$v$ gives the conserved canonical momentum
$\partial L/\partial\dot{v}=m\Omega^2\dot v+\beta u^2=\ell=const$. The Legendre transform at
fixed $\ell$ separates the transverse kinetic and potential terms,
\begin{equation}\label{Hclass2}
	H=\frac{p_u^2}{2M(u)}+V_{\rm eff}(u),
\end{equation}
where $M(u)=m\Omega^2(u)>0$ is the transverse inertial coefficient.
The longitudinal kinetic energy, expressed using the mechanical
momentum $\ell-\beta u^2$, supplies the effective potential
\begin{equation}\label{Veff}
	V_{\rm eff}(u)=\frac{(\ell-\beta u^2)^2}{2m\Omega^2(u)}.
\end{equation}
The reduced equation of motion is
$M\ddot u+\tfrac12 M'\dot u^2=-V_{\rm eff}'(u)$. For a given $u(t)$, integration of $\dot v=(\ell-\beta u^2)/(m\Omega^2)$
provides the longitudinal motion. A periodic transverse orbit can have a net longitudinal drift.

The conserved energy $H=E$ results in $\dot u^2=2(E-V_{\rm eff})/M$, so
turning points, where $\dot{u}=0$, satisfy $V_{\rm eff}=E$. For the ordinary helicoid,
the transverse coefficient is $m$ whereas in the current case it is $M(u)$. However, the longitudinal metric
and gauge potential provide the same $V_{\rm eff}$~\cite{Guvendi2026}.
On an allowed monotonic segment of the conformal surface, the
positive time element is
\begin{equation}\label{eq:period-comparison}
	\mathrm{d}t=\Omega(u)\sqrt{\frac{m}{2[E-V_{\rm eff}(u)]}}\,
	|\mathrm{d}u|.
\end{equation}
For the ordinary-helicoid the integrand is sole the square-root factor.
For identical parameters $E,\ell,qB,\omega$, both systems have the same allowed
intervals. Since $\Omega>1$ at $u\ne0$, every nontrivial periodic
transverse motion takes longer on the conformal surface.
For $qB\ne0$, $V_{\rm eff}$ grows as $q^2B^2u^2/(8m)$ at large
$|u|$, so finite-energy motion is transversely confined.

\subsection{Equilibria, Turning Points, and Phase Diagram}

\begin{figure}[!htbp]
	\centering
	\includegraphics[width=\columnwidth]{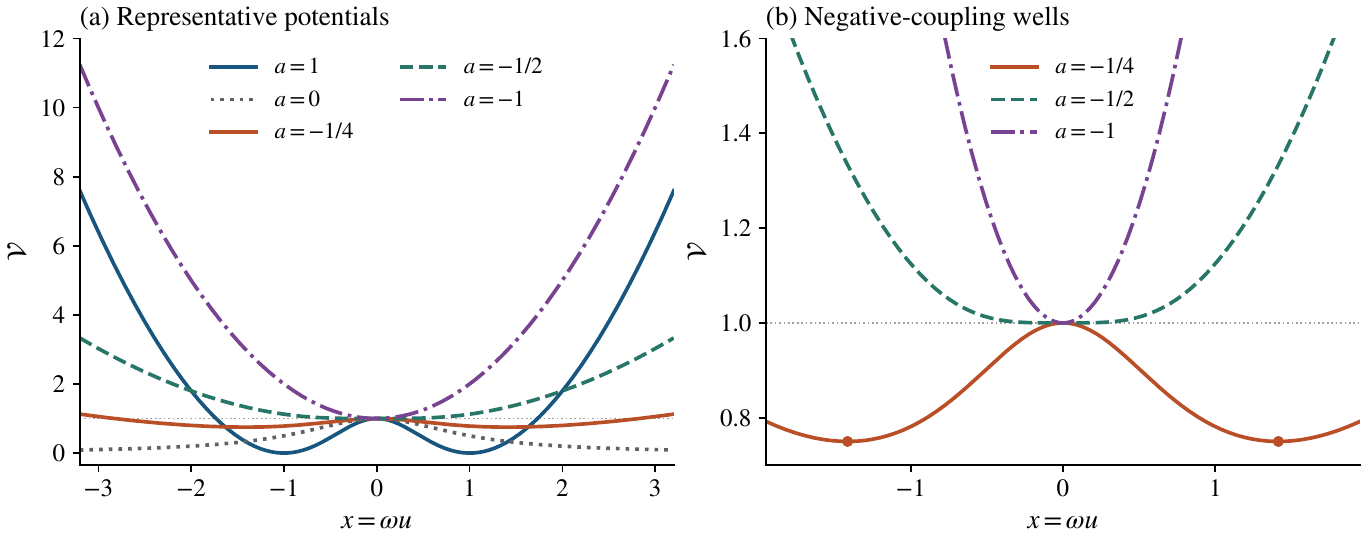}
	\caption{Scaled effective potential for representative $a$.
	(a) The positive-$a$ double well, the zero-field barrier, and
	three negative-$a$ cases. (b) Negative-$a$ regimes near the center.
	At $a=-1/4$ the minima are at $x=\pm\sqrt{2}$ with
	$\mathcal V=3/4$. At $a=-1/2$ the central minimum is quartic,
	and at $a=-1$ it is quadratic. The dotted horizontal line marks
	$\mathcal V=1$.}
	\label{fig:Figure1}
\end{figure}

For $\ell\ne0$, let us introduce the dimensionless quantities $x=\omega u$, $P=p_u/\ell$,
$a=\beta/(\ell\omega^2)$, and $\varepsilon=2mE/\ell^2$. The scaled dimensionless
Hamiltonian then +is
\begin{equation}\label{eq:scaled-h}
	h(x,P)\equiv\frac{2mH}{\ell^2}=
	\frac{P^2+(1-ax^2)^2}{1+x^2}.
\end{equation}
Let $\mathcal V(x)=h(x,0)=\mathcal V(-x)$ denote the scaled effective potential.
Its derivative factors as
\begin{equation}\label{eq:potential-derivative}
	\mathcal V'(x)=
	-\frac{2x(1-ax^2)(1+2a+ax^2)}{(1+x^2)^2}.
\end{equation}
For $a>0$, the two minima at $x=\pm a^{-1/2}$ have zero energy
and $\ell-\beta u^2=0$, so the particle is at rest there.
For $-1/2<a<0$, the minima are at $x=\pm\sqrt{-(1+2a)/a}$
with $\varepsilon_{\min}=-4a(1+a)>0$. In both ranges the origin
is a maximum at $\varepsilon=1$.
Near the origin the effective potential behaves like
$\mathcal V(x)=1-(1+2a)x^2+(1+a)^2x^4+O(x^6)$.
At $a=-1/2$, the two minima merge into a quartic minimum with
$\mathcal V(x)=1+x^4/[4(1+x^2)]$. For $a<-1/2$ the origin is a
quadratic minimum. In the particular case $a=-1$,
$\ell=-\beta/\omega^2$, so $\dot v=\ell/m$ along every orbit
and $\mathcal V(x)=1+x^2$. At $a=0$, $\mathcal V(x)=1/(1+x^2)$
is the zero-field barrier. Figure~\ref{fig:Figure1} illustrates
these changes.

Turning points satisfy $\mathcal V=\varepsilon$. With $\rho=x^2$
this gives $a^2\rho^2-(2a+\varepsilon)\rho+1-\varepsilon=0$.
For $a\ne0$ the algebraic roots are
\begin{equation}\label{rhodef}
	\rho_\pm=\frac{2a+\varepsilon
	\pm\sqrt{\varepsilon^2+4a(1+a)\varepsilon}}{2a^2}.
\end{equation}
Each real, nonnegative root gives $x=\pm\sqrt{\rho_\pm}$.
For $\varepsilon_{\min}<\varepsilon<1$ in either double well,
both roots are positive and the motion is in one of two disjoint
regions $x\in[-\sqrt{\rho_+},-\sqrt{\rho_-}]$ or $x\in[\sqrt{\rho_-},\sqrt{\rho_+}]$.
At the separatrix, $E=E_{\rm sep}=\ell^2/(2m)$, we have $\varepsilon =1$ and the inner root $\rho_-$ vanishes. The outer root is $\rho_+=(1+2a)/a^2$. The reduced phase-space contour consists of two
homoclinic loops meeting at the saddle. Above the barrier only
$\rho_+$ is positive, and the orbit crosses the axis between
$-\sqrt{\rho_+}$ and $\sqrt{\rho_+}$.
Table~\ref{tab:phase-diagram} summarises these findings and in addition contains the single-well and
zero-field regimes.

\begin{table}[!htbp]
	\centering
	\small
	\caption{Transverse phase diagram for $\ell\ne0$.
	In each confining well, equality at the minimum energy gives
	an equilibrium of the transverse dynamics.\\ }
	\label{tab:phase-diagram}
	\begin{tabular}{@{}p{0.16\textwidth}p{0.34\textwidth}p{0.41\textwidth}@{}}
		\toprule
		Range of $a$ & Stationary structure & Nonstationary transverse motion \\
		\midrule
		$a>0$ & Double well, $\varepsilon_{\min}=0$ &
			Two periodic components for $0<\varepsilon<1$, homoclinic
			loops at $\varepsilon=1$, and one component for $\varepsilon>1$. \\
		\addlinespace[2pt]
		$-1/2<a<0$ & Double well, $\varepsilon_{\min}=-4a(1+a)$ &
			Two periodic components for $\varepsilon_{\min}<\varepsilon<1$,
			homoclinic loops at $\varepsilon=1$, and one component for $\varepsilon>1$. \\
		\addlinespace[2pt]
		$a=-1/2$ & Quartic minimum, $\varepsilon_{\min}=1$ &
			One periodic component for $\varepsilon>1$. \\
		\addlinespace[2pt]
		$a<-1/2$ & Quadratic minimum, $\varepsilon_{\min}=1$ &
			One periodic component for $\varepsilon>1$. \\
		\addlinespace[2pt]
		$a=0$ & Barrier with maximum $\mathcal V(0)=1$ &
			Reflection for $0<\varepsilon<1$, critical branches at $\varepsilon=1$,
			and axis crossing for $\varepsilon>1$. Each trajectory is
			unbounded in at least one time direction. \\
		\bottomrule
	\end{tabular}
\end{table}

\begin{figure}[!htbp]
	\centering
	\includegraphics[width=\columnwidth]{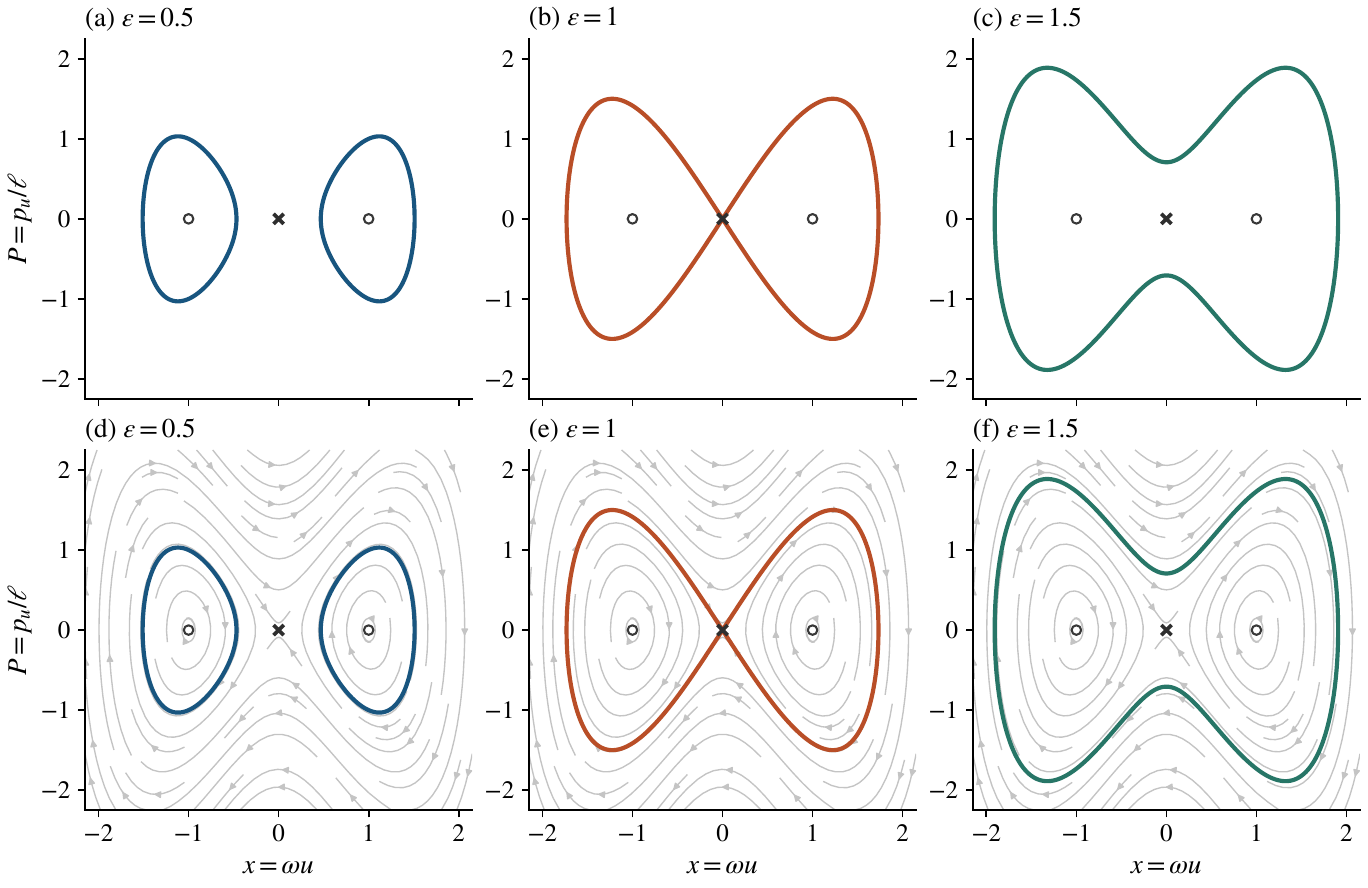}
	\caption{Reduced phase-space contours for $a=1$ at
	$\varepsilon=0.5$, $1$, and $1.5$ in panels (a)--(c), with
	the corresponding flow in panels (d)--(f). The bullets mark
	the stable equilibria and the cross marks the saddle point. Both
	homoclinic loops approach the saddle at $\varepsilon=1$.}
	\label{fig:3}
\end{figure}

The scaled orbit variables obey the Poisson bracket relation $\{x,P\}=\omega/\ell$.
With $\tau=(\ell\omega/m)t$, their evolution is generated by
$h/2$ under the unit bracket. Figure~\ref{fig:3} shows three
energy contours and the corresponding flows for $a=1$.

\FloatBarrier

So far our discussion was for the case $\ell\neq 0$.
At $\ell=0$ the scaled energy above is undefined, so
Eq.~\eqref{Veff} supplies the classification directly. For
$qB\ne0$ this potential is $\beta^2u^4/[2m\Omega^2(u)]$,
with a quartic minimum at the origin and one axis-crossing
oscillatory interval at every $E>0$.

In the zero-field limit the effective potential is
$E_{\rm sep}/\Omega^2(u)$.
For $E>0$, let $d=1-E_{\rm sep}/E$. On each monotonic
branch Eq.~\eqref{eq:period-comparison} gives
\begin{equation}\label{eq:zero-field-time}
	t-t_i=\pm\sqrt{\frac{m}{2E\omega^2}}\,
	\bigl[I_d(x)-I_d(x_i)\bigr],
\end{equation}
where $I_d'(x)=(1+x^2)/\sqrt{x^2+d}$ on the allowed domain.
For $d>0$, the orbit can cross the axis. The integral is explicitly given by
\begin{equation}\label{eq:zero-field-above}
	I_d(x)=\tfrac12 x\sqrt{x^2+d}+
	(1-d/2)\operatorname{arsinh}\!\left(\frac{x}{\sqrt d}\right).
\end{equation}
This expression includes the case $\ell=0$ through $d=1$,
for which $v$ is constant. If $E=\ell=0$ at $B=0$, the
particle is at rest.

For $d=-\alpha<0$, a branch reflects at $|x|=\sqrt{\alpha}$
and stays on the same side of the axis. A primitive valid on
either side is
\begin{equation}\label{eq:zero-field-below}
	I_{-\alpha}(x)=\tfrac12 x\sqrt{x^2-\alpha}
	+\operatorname{sgn}(x)(1+\alpha/2)
	\operatorname{arcosh}\!\left(\frac{|x|}{\sqrt{\alpha}}\right).
\end{equation}
At $E=E_{\rm sep}>0$, the nonstationary branches use
$I_0(x)=\operatorname{sgn}(x)(\ln|x|+x^2/2)$. They approach the
central saddle in one time direction and escape in the other.
In the escaping direction, the large-$|x|$ quadratic term in
the time integral gives
$|u(t)|\sim[8E/(m\omega^2)]^{1/4}|t|^{1/2}$.

\subsection{Periods, Separatrix, and Weak-Twist Limit}\label{sec:periods}

\begin{figure}[!t]
	\centering
	\includegraphics[width=\textwidth]{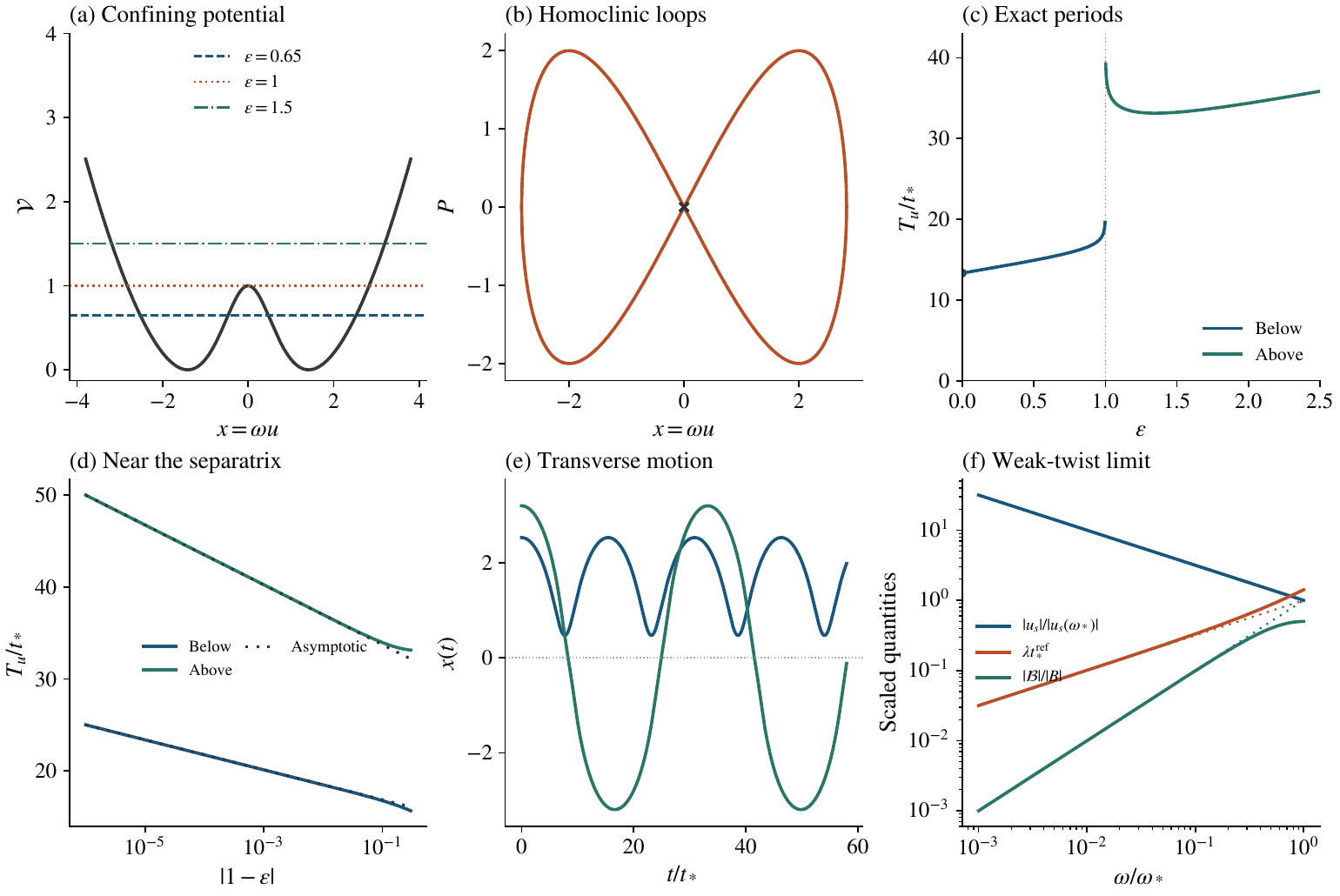}
	\caption{Transverse motion and periods for $a=1/2$ and $\ell>0$
	in panels (a)--(e), with time in units of $t_*=m/(|\ell|\omega)$.
	(a) Effective potential and three energy levels. (b) Homoclinic
	loops at $\varepsilon=1$. (c) Exact periods, with the marker at
	the small-amplitude limit. (d) Exact periods and dotted asymptotic
	curves from Eqs.~\eqref{eq:period-asym-below} and
	\eqref{eq:period-asym-above}, including their finite terms.
	(e) Intrawell and axis-crossing orbits at $\varepsilon=0.65$
	and $1.5$, respectively, in the colors used in (a).
	(f) Weak-twist behavior at fixed $qB=\ell\omega_*$, with
	$\omega_*>0$, $t_*^{\rm ref}=m/(|\ell|\omega_*)$, and the field
	evaluated at $u=1/\omega_*$. The dotted lines show the limiting
	power laws for the instability rate and the field.}
	\label{fig:4}
\end{figure}

Let $t_*=m/(|\ell|\omega)$. The time element in the scaled variables
is $dt_*=(1+x^2)|\mathrm{d}x|/|P|$. In either double well and below
the barrier, $P^2=a^2(\rho_+-x^2)(x^2-\rho_-)$.
Integrating over a full oscillation in one well using
$x^2=\rho_+(1-k^2\sin^2\theta)$ gives
\begin{equation}\label{eq:period-below}
	T_{<}=\frac{2t_*}{|a|}
	\left[\frac{\mathsf K(k)}{\sqrt{\rho_+}}
	+\sqrt{\rho_+}\,\mathsf E(k)\right].
\end{equation}
Here we have set $k^2=1-\rho_-/\rho_+$. $\mathsf K$ and $\mathsf E$ are the complete elliptic integrals with modulus $k$ of first and second kind, respectively.
Above the barrier, let $b=\sqrt{\rho_+}$ and $\gamma=-\rho_->0$.
The momentum factors as $P^2=a^2(b^2-x^2)(x^2+\gamma)$,
and a full cycle crosses the axis twice. Substitution
$x=b\sin\theta$ on the positive half gives
\begin{equation}\label{eq:period-above}
	T_{>}=\frac{4t_*}{|a|}
	\frac{(1-\gamma)\mathsf K(k)+(b^2+\gamma)\mathsf E(k)}
	{\sqrt{b^2+\gamma}}.
\end{equation}
In Eq.~\eqref{eq:period-above} the modulus is $k^2=b^2/(b^2+\gamma)$.
Both expressions are finite away from the separatrix in their
respective periodic regimes.

At a non-degenerate minimum $u_{\rm s}$ we have harmonic motion with a frequency following from
linearizing the reduced equation of motion,
\begin{equation}\label{Omegaosc}
	\omega_{\rm s}^2=\frac{V_{\rm eff}''(u_{\rm s})}{M(u_{\rm s})}.
\end{equation}
For $a>0$ this gives $\omega_{\rm s}=2|\ell|\omega a^{3/2}/[m(1+a)]$,
which equals $|q\mathcal B(u_{\rm s})|/m$.
For $-1/2<a<0$, it gives
$\omega_{\rm s}=2|\ell|\omega|a|^{3/2}\sqrt{1+2a}/[m(1+a)]$.
In both cases Eq.~\eqref{eq:period-below} tends to
$2\pi/\omega_{\rm s}$ as the two turning points of one well
coalesce at its minimum.

In the double-well sectors put $\eta=1+2a>0$. The central
curvature of the potential is $V_{\rm eff}''(0)=-\ell^2\omega^2\eta/m$.
With $M(0)=m$, the linearized dynamics is $\ddot u=\lambda^2u$,
with instability rate
\begin{equation}\label{eq:lambda}
	\lambda=\frac{\sqrt{\eta}}{t_*}.
\end{equation}
The ordinary helicoid has the same inertia at the axis and the same
rate. On a homoclinic loop, $P^2=x^2(\eta-a^2x^2)$, and the time
element near the saddle is $|\mathrm{d}x|/(\lambda|x|)$.

For fixed $a$ in either double-well range, the roots near
$\varepsilon=1$ satisfy $|\rho_-|=\delta_\pm/\eta+O(\delta_\pm^2)$
and $\rho_+=\eta/a^2+O(\delta_\pm)$, with
$\delta_-=1-\varepsilon$ below and $\delta_+=\varepsilon-1$ above.
On both sides the squared complementary modulus
$k'^2=1-k^2$ is $a^2\delta_\pm/\eta^2+O(\delta_\pm^2)$.
Using $\mathsf K(k)=\ln(4/k')+O(k'^2|\ln k'|)$ and
$\mathsf E(k)=1+O(k'^2|\ln k'|)$~\cite[Sec.~19.12]{DLMF}, the
single-well period becomes
\begin{equation}\label{eq:period-asym-below}
	T_{<}=\frac{1}{\lambda}
	\left[\ln\!\left(\frac{16\eta^2}{a^2\delta_-}\right)
	+\frac{2\eta}{a^2}\right]+O(t_*\delta_-|\ln\delta_-|).
\end{equation}
An orbit above the barrier passes through the slow region near the
saddle twice in a cycle, giving
\begin{equation}\label{eq:period-asym-above}
	T_{>}=\frac{2}{\lambda}
	\left[\ln\!\left(\frac{16\eta^2}{a^2\delta_+}\right)
	+\frac{2\eta}{a^2}\right]+O(t_*\delta_+|\ln\delta_+|).
\end{equation}

Figure~\ref{fig:4} illustrates in panels (a), (b) and (e) the potential, the homoclinic loop and representative trajectories in a double well. The panels (c) and (d) compare the exact periods with the separatrix expansions. Panel (f) explicates the weak-twist limit.

In the weak-twist limit at fixed $B$ and $\ell$, the longitudinal
first integral gives $\dot v=\ell/m-qB\omega u^2/(2m)+O(\omega^2)$
at fixed $u$. Over bounded intervals of $u$ and velocity, expansion
of the transverse equation gives
\begin{equation}\label{eq:weak-acceleration}
	\ddot u=\frac{qB\ell\omega}{m^2}u+O(\omega^2).
\end{equation}
The leading local correction is hyperbolic for $qB\ell>0$ and
oscillatory for $qB\ell<0$. In either case its inverse time scale
vanishes as $\sqrt{\omega}$. With fixed finite initial data the
limiting motion over finite time is free on the planar embedding,
with $u(t)\to u_i+\dot u_i(t-t_i)$ and
$v(t)\to v_i+\ell(t-t_i)/m$.

The limit is nonuniform in $u$. For $qB\ell>0$, the minima move out
with $|u_{\rm s}|=\sqrt{2|\ell|/(|qB|\omega)}$ and
$\lambda\sim\sqrt{qB\ell\omega}/m$, while the field extrema move
out as $1/\omega$. Here $a=qB/(2\ell\omega)$ diverges as the twist
rate decreases. Holding $a$ fixed would require $qB/\ell$ to scale
with $\omega$.

\FloatBarrier

\section{Quantum Hamiltonian and Semiclassical Spectrum}\label{sec:semiclassical}

We quantize a spinless particle intrinsically using
$\widehat H=-\hbar^2\Delta_A/(2m)$, where $\Delta_A$ is the magnetic
Laplace--Beltrami operator. This choice sets the additional scalar
curvature coupling to zero~\cite{DeWitt1952}. A thin-layer model
with physical normal confinement can induce potentials from the
extrinsic geometry and the normal states~\cite{daCosta1981,FerrariCuoghi2008,SchusterJaffe2003}.
At fixed $\ell=\hbar k$, the separation below in Sec.~\ref{sec:quantum}
gives the transverse operator
\begin{equation}\label{eq:fiber-operator}
	H_u=-\frac{\hbar^2}{2M(u)}\frac{\mathrm{d}^2}{\mathrm{d}u^2}
	+V_{\rm eff}(u).
\end{equation}
The inner product uses the weight $\Omega^2(u)$. Integration
by parts for smooth compactly supported functions gives
\begin{equation}\label{eq:quadratic-form}
	\langle f,H_u g\rangle_{\Omega^2}
	=\frac{1}{2m}\int_{\mathbb R}
	\left[\hbar^2\overline{f'}g'
	+(\ell-\beta u^2)^2\overline f g\right]\,\mathrm{d}u.
\end{equation}
The closure of this nonnegative form defines the self-adjoint
transverse Hamiltonian.

For the semiclassical analysis, let $s(u)=\int_0^u\Omega(y)\,\mathrm{d}y$
be transverse arclength and set $F(s)=\ln\Omega(u(s))$.
Since $\Omega^2\,\mathrm{d}u=\Omega\,\mathrm{d}s$, the map
$\psi_k(u)\mapsto\Omega(u(s))^{1/2}\psi_k(u(s))$ is unitary
to the measure $\mathrm{d}s$. In that measure Eq.~\eqref{eq:fiber-operator}
becomes
\begin{equation}\label{eq:arclength-operator}
	\widetilde H_u=-\frac{\hbar^2}{2m}\frac{\mathrm{d}^2}{\mathrm{d}s^2}
	+V_{\rm eff}(u(s))+\frac{\hbar^2}{2m}
	\left(\tfrac12 F''(s)+\tfrac14 F'(s)^2\right).
\end{equation}
The last term comes from the unitary change of variables and is
of order $\hbar^2$. We will ignored it in our semiclassical analysis. The leading oscillation frequency at a
nondegenerate minimum is therefore the classical
$\omega_{\rm s}$ in Eq.~\eqref{Omegaosc}. The local harmonic
estimate for eigenvalues is
\begin{equation}\label{eq:harmonic-general}
	E_n^{\rm harm}=V_{\rm eff}(u_{\rm s})
	+\hbar\omega_{\rm s}\left(n+\tfrac12\right).
\end{equation}
The offset is zero for $a>0$ and equals
$-4a(1+a)\ell^2/(2m)$ for $-1/2<a<0$. For $a<-1/2$,
the offset at the central minimum is $\ell^2/(2m)$.
Introducing the cyclotron frequency  $\omega_c=|qB|/m$, the central frequency obeys
\begin{equation}\label{eq:central-frequency}
	\Omega_0^2=\frac{|\ell|\omega}{m}
	\left(\omega_c-\frac{|\ell|\omega}{m}\right).
\end{equation}
This is $V_{\rm eff}''(0)/m$ in the central-well regime.
The quartic minima at $a=-1/2$ and at $\ell=0$, $qB\ne0$
require an anharmonic treatment.

On a classically allowed branch, the magnitude of the
transverse momentum is $p(u;E,\ell)=\sqrt{2M(u)[E-V_{\rm eff}(u)]}$.
The action is independent of the choice of transverse coordinate
because $p_s\,\mathrm{d}s=p_u\,\mathrm{d}u$. Between two
simple turning points $u_{\rm L}<u_{\rm R}$, leading-order WKB
quantization gives
\begin{equation}\label{eq:bohr-sommerfeld}
	I(E_n^{(0)})=\pi\hbar\left(n+\tfrac12\right),
\end{equation}
where $I(E)=\int_{u_{\rm L}}^{u_{\rm R}}p(u;E,\ell)\,\mathrm{d}u$.
For $a\ne0$ at $\ell\ne0$, set $r_\pm=\rho_\pm/\omega^2$ using
Eq.~\eqref{rhodef}. In the right well below the barrier, the
limits are $\sqrt{r_-}$ and $\sqrt{r_+}$. Above the barrier
the limits are $-\sqrt{r_+}$ and $\sqrt{r_+}$. Hence, \eqref{eq:bohr-sommerfeld} may be applicable for both case but cannot be used for the tunneling splitting.

In a double well with $V_{\rm eff}(u_{\rm s})<E<E_{\rm sep}$,
tunneling across the central forbidden interval splits the
even and odd energy levels. The barrier action is
\begin{equation}\label{eq:tunnel-action}
	S_{\rm tun}(E)=\int_{-\sqrt{r_-}}^{\sqrt{r_-}}
	\sqrt{2M(u)[V_{\rm eff}(u)-E]}\,\mathrm{d}u.
\end{equation}
In the positive-$a$ well-bottom limit, the integral reduces to
$S_{\rm tun}(0)=4|\ell|\,|u_{\rm s}|/3$. At matched energy and
longitudinal momentum, the action integrand of the conformal
helicoid is the ordinary-helicoid integrand times $\Omega(u)$.

For a semiclassically excited doublet, with isolated one-well
level $E_n^{(0)}$ and classical period $T_{<}=2\,\mathrm{d}I/\mathrm{d}E$,
the leading splitting is
\begin{equation}\label{eq:wkb-splitting}
	\Delta E_n\simeq\frac{2\hbar}{T_{<}(E_n^{(0)})}\,
	\exp[-S_{\rm tun}(E_n^{(0)})/\hbar].
\end{equation}
At fixed low $n$, matching to the harmonic wave function
gives a different prefactor~\cite[App.~B]{Garg2000}.
With $\nu=n+\tfrac12$ and $g_n=\sqrt{2\pi}\,\nu^\nu e^{-\nu}/n!$,
it yields
\begin{equation}\label{eq:tunnel-splitting}
	\Delta E_n\simeq g_n\frac{\hbar\omega_{\rm s}}{\pi}\,
	\exp[-S_{\rm tun}(E_n^{\rm harm})/\hbar].
\end{equation}
Both formulas require $S_{\rm tun}\gg\hbar$ at the relevant level.
The low-$n$ expression also assumes a small oscillator energy
above the well minimum compared with the barrier height.
Near the separatrix the merging turning points require uniform
matching.

\section{Quantum Dynamics on the Conformal Helicoidal Surface}\label{sec:quantum}

Minimal coupling of a vector potential requires the covariant derivative given by $D_i=\partial_i-iqA_i/\hbar$ and the associate magnetic Laplace--Beltrami operator reads $\Delta_A\psi=g^{-1/2}D_i(\sqrt g\,g^{ij}D_j\psi)$.
For the metric \eqref{ds2}, $\sqrt g\,g^{ij}=\delta^{ij}$.
The gauge in Eq.~\eqref{Av} then gives
\begin{equation}\label{eq:quantum-H}
	\widehat H=-\frac{\hbar^2}{2m\Omega^2(u)}
	\left[\partial_u^2+
	\left(\partial_v-\frac{i\beta u^2}{\hbar}\right)^2\right].
\end{equation}
The full inner product uses the measure
$\Omega^2(u)\,\mathrm{d}u\,\mathrm{d}v$.
The coefficients are independent of $v$ and time, so we use the ansatz
$\Psi(u,v,t)=e^{ikv-iEt/\hbar}\psi_k(u)$.
Substituting this in the stationary equation and multiplying
by $2m\Omega^2$ gives
\begin{equation}\label{eq:weighted-dimensional}
	-\hbar^2\psi_k''(u)+(\hbar k-\beta u^2)^2\psi_k(u)
	=2mE\Omega^2(u)\psi_k(u).
\end{equation}
Here $\hbar k$ is the canonical momentum and $\hbar k-\beta u^2$
is the mechanical momentum associated with the $v$-coordinate. A transverse bound state obeys
$\int_{\mathbb R}\Omega^2|\psi_k|^2\,\mathrm{d}u<\infty$.
For $qB\ne0$, the potential in Eq.~\eqref{eq:fiber-operator}
diverges at both ends, so the form \eqref{eq:quadratic-form}
controls the transverse tails. Local Sobolev compactness on
bounded intervals then gives a discrete spectrum at each fixed
$k$. The full surface problem is a Fourier superposition over
continuous $k$, since $v\in\mathbb R$.

Let $\kappa=k/\omega$, $b_{\rm q}=qB/(\hbar\omega)$, and
$\mathcal E=2mE/(\hbar^2\omega^2)$. With $x=\omega u$,
Eq.~\eqref{eq:weighted-dimensional} becomes the weighted
eigenvalue problem
\begin{equation}\label{eq:weighted-dimensionless}
	\left[-\frac{\mathrm{d}^2}{\mathrm{d}x^2}
	+\left(\kappa-\frac{b_{\rm q}x^2}{2}\right)^2\right]\psi
	=\mathcal E(1+x^2)\psi .
\end{equation}
At $\kappa\ne0$, the classical parameters are
$a=b_{\rm q}/(2\kappa)$ and $\varepsilon=\mathcal E/\kappa^2$.
For a numerical check, we use $b_{\rm q}=8$ and
compare a double well at $\kappa=8$ with a central well at $\kappa=-4$.
A centered finite-difference discretization with Dirichlet
endpoints provides the levels in Tables~\ref{tab:double-spectrum}
and~\ref{tab:central-spectrum}. Grid refinement and domain
enlargement change the extrapolated eigenvalues by less than
$3\times10^{-9}$ in $\mathcal E$.

\begin{table}[!htbp]
	\centering
	\small
	\caption{Double-well levels at $(\kappa,b_{\rm q})=(8,8)$.
	The index $n$ denotes the local oscillator level and
	$\Delta\mathcal E$ is the odd level minus the even level.\\}
	\label{tab:double-spectrum}
	\begin{tabular}{@{}crrrrr@{}}
		\toprule
		$n$ & Harmonic & Even & Odd & $\Delta\mathcal E$ & WKB splitting \\
		\midrule
		0 & 3.771236 & 3.72439537 & 3.72440133 & $5.97\times10^{-6}$ & $6.12\times10^{-6}$ \\
		1 & 11.313708 & 11.06922524 & 11.06930420 & $7.90\times10^{-5}$ & $8.81\times10^{-5}$ \\
		\bottomrule
	\end{tabular}\par\medskip
	\caption{Central-well levels at $(\kappa,b_{\rm q})=(-4,8)$.
	The harmonic energies include the offset $\mathcal E_{\min}=16$.\\}
	\label{tab:central-spectrum}
	\begin{tabular}{@{}crr@{}}
		\toprule
		$n$ & Harmonic & Numerical \\
		\midrule
		0 & 20.000000 & 20.13828345 \\
		1 & 28.000000 & 27.21117982 \\
		2 & 36.000000 & 33.85885476 \\
		\bottomrule
	\end{tabular}
\end{table}

\FloatBarrier

For the double well, the dimensionless ground-doublet splitting
is $5.97\times10^{-6}$, and Eq.~\eqref{eq:tunnel-splitting}
gives $6.12\times10^{-6}$. The WKB result for the next doublet is
about $12\%$ larger than the numerical splitting.
In the central well the harmonic estimate is within $1\%$ of the
ground-state energy, with larger deviations for the excited states.

\section{Summary and Discussion}
\addcontentsline{toc}{section}{Summary and Discussion}
Embedding the helicoid in four-dimensional Euclidean space changes the ordinary helicoid into an conformal helicoid. Whereas the longitudinal metric coefficient and the pulled-back gauge potential are preserved, the transverse inertia and the
surface area element are changed.
For the same energy parameter the classical turning points are identical but the larger transverse inertia increases the period of every nontrivial transverse oscillation on the conformal surface. The
intrinsic magnetic field there reaches a magnitude of $|B|/2$ at the zeros of the
Gaussian curvature and decays at both transverse ends.

For nonzero magnetic coupling the effective potential grows quadratically
at large $|u|$, confining any finite-energy transverse motion. The two minima
of the negative-$a$ double well merge into a quartic minimum at the origin when $a$ approaches the value $a=-1/2$.
In both double-well regimes, the separatrix divides oscillations confined
to one well from axis-crossing motion. The exact periods diverge
logarithmically as the energy approaches the barrier, with twice the
leading coefficient for an axis-crossing cycle. Both helicoids share this
leading divergence because their transverse inertias agree at the saddle.
Removing the magnetic field permits transverse escape. The weak-twist
limit gives free motion on bounded patches over finite times. The well
minima for $qB\ell>0$ move outward as $\omega^{-1/2}$, so the limit is
nonuniform across the surface.

Intrinsic quantization gives a weighted eigenvalue problem with discrete
transverse levels at each fixed longitudinal momentum for $qB\ne0$.
At matched energy and momentum the conformal metric increases the
double-well barrier action. In Table~\ref{tab:double-spectrum}, the
dimensionless ground-doublet splitting is $5.97\times10^{-6}$, and the
low-level WKB formula overestimates the first two splittings by about
$3\%$ and $12\%$, respectively. Measured from the central-well minimum, the harmonic
ground-state energy is about $3.3\%$ below the numerical value. The two
excited levels in Table~\ref{tab:central-spectrum} show larger deviations.

As an alternative to the present quantization approach, which is based on local Schrödinger-like operators, one may use Feynman's path integral formalism. Path integral quantization, in general, is more difficult compared to the operator approach. However, it may provide additional insides when non-trivial topologies are involved. It may even be possible to do an exact path integration as shown in recent works on conical space and spaces with topological defects \cite{Inomata2012a,Inomata2012b}.

\section*{References}

\begingroup
\renewcommand{\section}[2]{}

\endgroup

\end{document}